\documentclass[twocolumn,prl,aps]{revtex4}
\usepackage{graphicx}                                   
\tighten
\renewcommand{\vec}[1]{{\mathbf{#1}}}

\newcommand{\beq}{\begin{eqnarray}}
\newcommand{\eeq}{\end{eqnarray}}
\begin{document}
\draft

\title
{Doped Mott Insulators are Insulators: Hole localization in the Cuprates}
\author{  Ting-Pong Choy and Philip Phillips}
\affiliation{Loomis Laboratory of Physics,
University of Illinois at Urbana-Champaign,
1110 W.Green St., Urbana, IL., 61801-3080}

\begin{abstract}
We demonstrate that a Mott insulator lightly doped with holes is still an insulator at low temperature even without disorder.  Hole localization obtains because the chemical potential lies in a pseudogap which has a vanishing density of states at zero temperature.  The energy scale for the pseudogap is set by the nearest-neighbour singlet-triplet
 splitting. As this energy scale vanishes if transitions, virtual or otherwise, to the upper Hubbard band are not permitted, the fundamental length scale 
in the pseudogap regime is the average distance between doubly occupied sites.
Consequently, the pseudogap is tied to the non-commutativity of the two limits
 $U\rightarrow\infty$ ($U$ the on-site Coulomb repulsion) and $L\rightarrow\infty$ (the system size). 
\end{abstract}

\maketitle

Hole doping a Mott insulator shifts\cite{sawatzky} the chemical potential from the middle of the charge gap
generated by the energy cost ($U$) for double occupancy to the top of the lower
Hubbard band.  Nominally, the density of states at the top of the lower
Hubbard band is non-zero. Consequently, doped Mott insulators
are expected to be conductors.  However, doped Mott insulators such as the high temperature cuprate superconductors are well known to possess
a pseudogap\cite{alloul,norman,timusk} at the Fermi energy below some characteristic temperature, $T^\ast$, that persists well into the superconducting dome.  While a dip in the density of states is not sufficient to destroy the simple picture that a metallic state obtains upon light hole doping, certainly a vanishing density of the states at the Fermi level 
would be.  The question arises: Does the density of states vanish at the chemical potential in the limit $T\rightarrow 0$ in the underdoped cuprates or in lightly doped Mott insulators in general?  The analysis presented here on the Hubbard model suggests the answer to this question is yes and hence lightly doped Mott insulators are, in fact, still insulators.

Experimental probes that shed light, either directly or indirectly, on the ultimate fate of the density of states at the chemical potential in doped Mott systems are of three types: transport, tunneling and angle-resolved photoemission (ARPES).  Early transport measurements on La$_{2-x}$Sr$_x$CuO$_{4+y}$ revealed\cite{birgeneau1,birgeneau2} that in the lightly-hole doped regime, the in-plane resistivity obeys the 3-d variable hopping form,
\beq\label{div}
\rho(T)\propto e^{(T_0/T)^\alpha}
\eeq
with $\alpha=1/4$\cite{birgeneau1} or diverges logarithmically
 as $\ln T_0/T$\cite{birgeneau2}.  
In both cases, if localization is due to disorder (and hence extrinsic to 
Mott physics), an externally applied magnetic field should couple to the orbital motion and yield a negative magnetoresistance.  While the magnetoresistance is negative, it is independent of the direction of the field\cite{birgeneau2}, indicating that the localization mechanism is intrinsic and arises solely from spin scattering.  In fact, extensive measurements\cite{boeb1,boeb2,boeb3,boeb4} over the last 10
years indicate that once superconductivity is destroyed by the application of a large magnetic field, only two electrically
distinct phases exist in the cuprates: 1) an insulator with
 a logarithmically diverging resistivity
of the form $\ln T_0/T$ throughout the pseudogap region,
 $x<x_{\rm c}$, and 2) a Fermi liquid metal
for $x>x_{\rm c}$, where $x_c\approx x_{\rm opt}$, the optimal doping level.   In the absence of a field, the most recent\cite{ando} in-plane transport data on untwinned crystals of YBa$_2$Cu$_3$O$_{6.35}$,
a composition right at the edge of the superconducting dome, corroborate the 
diverging $\ln(T_0/T)$ behaviour found in the high magnetic field limit 
for both $\rho_a$ and $\rho_b$. These authors\cite{ando} conclude that the localization
mechanism is independent of field and likely to be a consequence of an
intrinsically insulating pseudogap at $T=0$.  Scanning tunneling experiments\cite{stm} are consistent with the deepening of the pseudogap as the temperature is decreased. Finally, recent ARPES measurements\cite{shen} have detected a finite gap over the entire Brillouin zone, including along the $d_{x^2-y^2}$ nodal line, in the {\bf normal} state of
La$_{2-x}$Sr$_x$CuO$_4$, Ca$_{2-x}$Na$_x$CuO$_2$Cl$_2$, and Nd$_{2-x}$Ce$_x$CuO$_4$.   In  Ca$_{2-x}$Na$_x$CuO$_2$Cl$_2$, the gap was observed to close at
$x=0.12$. Similarly, in optimally doped Bi2212\cite{campu}, the
 imaginary part of the self energy is momentum dependent but it remains non-zero even along the nodal directions. The ubiquity of a complete gap in the normal state of both electron and hole-doped cuprates prompted Shen, et. al.\cite{shen} to conclude that gapped excitations in lightly doped Mott insulators is a generic feature.

Quite generally, a pseudogap\cite{efros,altshuler} is an example of an orthogonality catastrophe\cite{mahan}. Typically, orthogonality leads to vanishing of both the quasiparticle weight, $Z$, as well as the conductivity at $T=0$.  In this case, we find that the orthogonality in a doped Mott insulator arises because hole transport is limited by the triplet-singlet energy gap.  The length scale underlying this energy gap is the average separation between doubly occupied
sites.  As this length scale diverges in projected models but remains finite in the Hubbard model, the pseudogap is tied to a non-commutativity
of $U\rightarrow\infty$ and $L\rightarrow\infty$.  This lack of commutativity
offers a possible explanation why all simulations thus far on the $t-J$ model find metallic transport\cite{haule,prelovsek} near half-filling whereas for the Hubbard model, an insulating state obtains.

\begin{figure}
\centering
\includegraphics[width=7.5cm,angle=270]{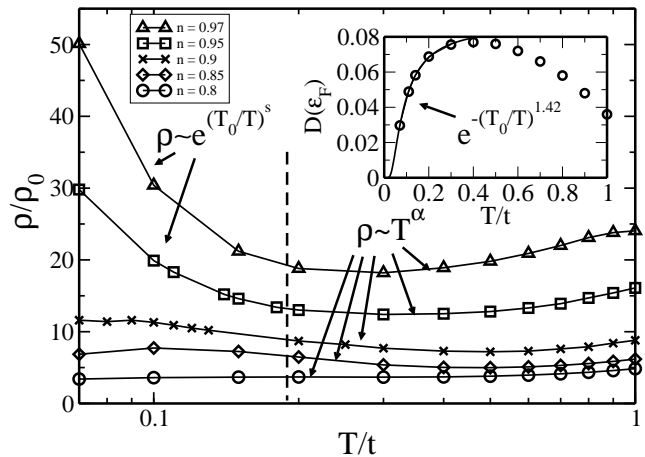}
\caption{Resistivity as a function of temperature calculated according to Eq. (\ref{cond}) for the Hubbard model (with $U=10t$) using the spectral function computed previously by Stanescu and Phillips\cite{fmott} for fillings $n=0.97, 0.95, 0.9, 0.85, 0.80$.  The inset indicates that the density of states at the chemical potential
vanishes as the temperature decreases giving rise to an insulating state
for $n\approx 1$.  $\rho_0=h/e^2$.}
\label{fig1} 
\end{figure}

The starting point for our analysis is the Hubbard model. In this context, we 
have been refining\cite{fmott} a non-perturbative resolvent method\cite{mm} for
calculating the 
single-particle spectral function
$A(\vec k,\omega)=-{\rm Im}FT(\theta(t-t')\langle \{c_{i\sigma}(t),c^\dagger_{j\sigma}(t')\}\rangle$
that is based on a self-consistent determination of the electron self-energy
using the Hubbard operators.  Here, $c_{i\sigma}$ is the electron
annihilation operator and FT represents the frequency and momentum Fourier transform.  In the spirit of cellular methods\cite{cell}, the essence of our 
procedure is to expand the electron self-energy for the 2D lattice
in terms of the resolvents for a small cluster. In our work, the eigenstates of a two-site cluster were used to expand the operators in the self-energy. As the self-energy can be written as a product of two operators, each of which can 
be centered on different lattice sites, a two-site expansion for each operator captures local correlations (albeit in a pair-wise fashion) over at most four lattice sites.  Such a local expansion has been shown to yield a heat capacity of the 1D Hubbard system\cite{fmott2} in excellent agreement with the Bethe ansatz as well as a pseudogap\cite{fmott} in the 2D Hubbard model. Our emphasis here is on using the spectral function to calculate the conductivity.  To obtain a direct link between the conductivity and the spectral function, we work with the non-crossing approximation 
\beq\label{cond}
{\mathop{\rm Re}\nolimits} \sigma _{xx} (0 + i\delta ) &=& 2
\pi e^2
\int d^2 k\int d\omega '(2t\sin k_x )^2 \nonumber\\
&& \left(  -\frac{\partial f(\omega ')}{\partial \omega '} \right)
\left[
A(\omega ',k) \right]^2
\eeq
to the Kubo formula for the
conductivity where $f(\omega)$ is the Fermi distribution function.  Although
Eq. (\ref{cond}) is only approximate, as it does not include vertex 
corrections, we will show that our conclusions are independent of any approximation used to compute the conductivity.  Shown in Fig. (\ref{fig1}) is the resultant computation of the resistivity as a function of temperature for fillings of $n=0.97$, $0.95$, $0.9$, $0.85$, and $n=0.8$.  At high temperatures the resistivity
increases algebraically regardless of the filling.  However, at low temperatues,  a divergence in accord with Eq. (\ref{div}) for fillings close to $n=1$ obtains.  We will determine the crossover filling on general grounds later.  In contrast, similar cluster treatments of the spectral function\cite{haule} of the t-J model coupled with Eq. (\ref{cond}) find a metallic conductivity at all fillings, even arbitrarily close to half-filling.  In fact, recent\cite{prelovsek} exact diagonalization calculations on finite clusters confirm the inherent metallic behaviour at low temperatures, regardless of filling, in the t-J model.
Metallic behaviour in the t-J model is consistent with the extensive numerical\cite{mischenko,sorella,dagotto} and self-consistent Born calculations\cite{born} which have found that a single hole is mobile in a quantum antiferromagnet described by the t-J model with a quasi-particle residue that scales as $Z\propto J/t$ where $J=4 t^2/U$.  

What then is the origin of the insulating state for the Hubbard model 
in the underdoped regime?   The inset in Fig. (\ref{fig1}) demonstrates that 
the density of states at the chemical potential plummets to zero exponentially as the temperature decreases.  The conductivity, Eq. (\ref{cond}), is a product of the derivative of the Fermi distribution
 function and the spectral function.  
Because the former is peaked while the latter is zero at the chemical potential, the product necessarily vanishes leading to an insulating state.  This cancellation persists to all orders of perturbation theory.  Hence, the insulating state found here is not an artifact of 
the approximate form of Eq. (\ref{cond}); rather it arises simply because $D(\epsilon_F)=0$ at $T=0$. 

\begin{figure}
\centering
\includegraphics[width=6.5cm,angle=270]{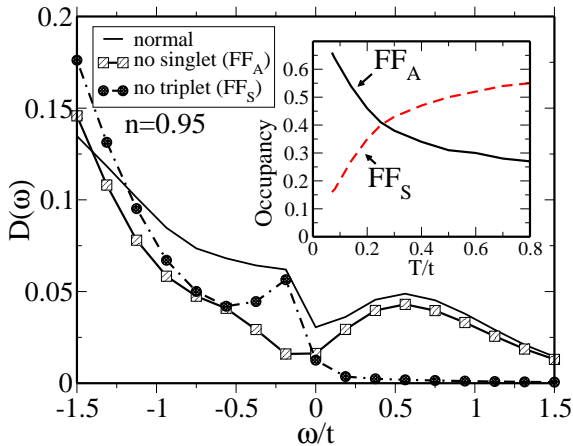}
\caption{Single-particle density of states when the contribution from various the nearest-neighbour singlet, $FF_A$, and triplet states, $FF_S$, are isolated. Elimination of the singlet contribution (solid circles) diminishes the density of states just below the chemical potential, whereas the triplet contribution (open squares) appears above the chemical potential. Since these states have an energy splitting of $J=4t^2/U$, their contributions to the density of states results in a real gap at $T=0$.The occupancy (see inset) in $FF_A$ and $FF_S$ crosses exactly at
$T^\ast$.}
\label{fig2}
\end{figure}

\begin{figure}
\centering
\includegraphics[width=6.5cm,angle=270]{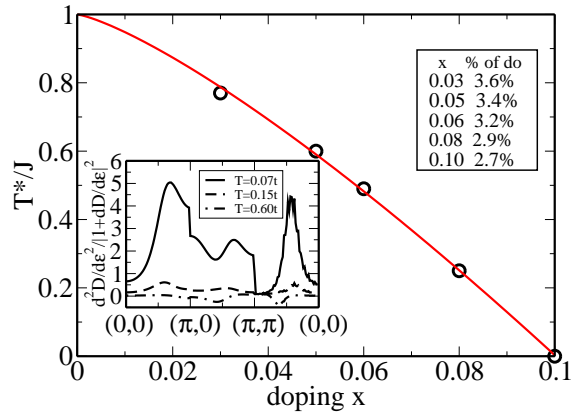}
\caption{Doping dependence of the pseudogap energy scale, $T^\ast$.  The dependence obeys the functional form, $J(1-x\xi_{\rm do}^2/4a^2)$, where $\xi_{\rm do}$ is the average distance between doubly occupied sites, the percentage of 
which is indicated in the table.  The inset shows the curvature of the density of states at the chemical potential as a function of momentum. Positive curvature at all momenta indicates that all momenta
contribute to the pseudogap.} 
\label{curv}
\end{figure}

Because the electron self-energy is expanded in the level operators\cite{fmott} for a two-site cluster, we can determine which local two-site correlations determine the physics of the vanishing 
of the density of states.  The solid line in Fig. (\ref{fig2}) illustrates clearly that the chemical potential lies in a local minimum in the single-particle density of states. This state of affairs obtains because nearest-neighbour singlet states (solid circles)
and triplet (open squares) contribute to the density of states just below and above the chemical potential, respectively as shown in Fig. (\ref{fig2}).  Because the triplet and singlet are split by an energy $J=4t^2/U$, their contributions to the density of states cannot occur at the same energy.  The density of states must have a dip which must constitute a real gap at $T=0$. The inset illustrates that precisely at the temperature (see Fig. (\ref{curv})) at which the dip in the density
of states obtains, the occupancy in the excited triplet states drops below that of the singlets.  This definitively proves that it is the singlet-triplet excitation gap that limits hole transport in a doped Mott insulator.
\noindent Such a pseudogap can be thought of as a spin gap\cite{alloul}
as in the context of a spin liquid\cite{rvb}.  Also consistent with our finding here is the ferromagnetic polaron 
picture\cite{manganites}.  However, neither experimental nor theoretical\cite{affleck} work supports the ferromagnetic polaron model in the parameter range of the cuprates.  In our simple picture that it is the local singlet-triplet splitting that gives rise to the pseudogap,  we expect the corresponding gap arising from the orthogonality to be isotropic in momentum space.  As illustrated by the inset in Fig. (\ref{curv}), the curvature of the density of states at the chemical potential is positive at each momentum indicating that all momenta contribute to the pseudogap, though with differing weights. This is consistent with the extensive ARPES study of Shen, et. al.\cite{shen}. In the context of the cuprates, we propose that any anisotropy\cite{appears} seen in the pseudogap is absent at $T^\ast$ but arises at lower temperatures as a result of any ordering phenomena\cite{yazdani} or pairing that might supervene on the 
pseudogap phase.  In fact, others\cite{jarrell} have concluded recently
based on cluster calculations on the Hubbard model that a pseudogap
arises entirely from local correlations independently of any ordering or pair formation. 

Two natural questions that arise from this work are 1) why do analogous cluster or exact diagonalization studies of the t-J model show no indication of localization\cite{haule, prelovsek} and 2) what sets the length scale for the energy gap. Both of these questions have the same answer.  Without the triplet contribution, the pseudogap in Fig. (\ref{fig2}) vanishes. However, the triplet contribution lies above the chemical potential and hence is part of the addition spectrum.
The addition spectrum\cite{sawatzky} of the low-energy spectral weight (LESW) is a sum of two distinct processes each involving spectral weight transfer between the upper and lower Hubbard bands: 1) a static part arising from state counting which grows as $2x$ but more importantly 2) a dynamical part that arises entirely from the hybridization. Since the triplet is present only when $t\ne 0$, the triplet contribution to the 
LESW is purely dynamical. In projected models in which double occupancy is eliminated at second order, the LESW scales exactly as $2x$\cite{sawatzky}. Hence, the dynamical contribution to the spectral weight transfer is absent.  However, the dynamical contribution to the addition part of the LESW can be treated perturbatively as first shown by Harris and Lange\cite{harris}.  Perturbation
theory alone is insufficient to generate a gap in an excitation spectrum
since the opening of a gap represents a phase transition.
 The essence of the problem is that as long as the insulating state is tied to the dynamical contribution to the spectral weight transfer between the upper and lower Hubbard bands, the length scale, $\xi_{\rm do}$, over which transport is governed by
 double occupancy must be finite.   That is,  the physics is sensitive to the order of limits of $U\rightarrow\infty$ and $L\rightarrow\infty$.  Such non-commutativity signals
a breakdown in perturbation theory as advocated previously\cite{phillips}.  
 $U\rightarrow\infty$, $L\rightarrow\infty$ results in $\xi_{\rm do}>L$, metallic transport. In the reverse order of limits, $\xi_{\rm do}<L$ and localization obtains provided that the $n_h\xi_{\rm do}^2<L^2$, $n_h=x(L/a)^2$ the number of holes. $n_h\xi_{\rm do}^2=L^2$ defines the percolation limit. By calculating the percentage of 
doubly occupied sites, we obtained $\xi_{\rm do}$ numerically and plotted the $T^\ast$-line, $J(1-cx(\xi_{\rm do}/a)^2)$, in Fig. (\ref{curv}). The agreement of this phenomenological fit with the resisitivity data in which a metallic state obtains at $x=0.1$ and the crossing in Fig. (\ref{fig2}) lend credence to our assertion that $\xi_{\rm do}$ is the relevant length scale for the pseudogap.
 Finally, the scaling form 
\beq
Z\propto L^{-(t/U)^p}\quad p>0
\eeq
for the one-hole quasiparticle weight lays plain that the discrepancy between the $t-J$\cite{mischenko,dagotto,born} and Hubbard\cite{sorella} results 
is one of lack of commutativity. In the $t-J$ model (no double occupancy), $U\rightarrow\infty$, $L\rightarrow\infty$ and $Z$ remains finite. In the reverse order of limits (Hubbard model), $Z$ vanishes.  

Indeed, other proposals for hole localization exist.  Some have argued that in the t-J model, a hole creates a phase string\cite{weng}.  However, such an exotic state is not borne out by extensive numerical
simulations on the t-J model\cite{mischenko}.  In the spin-fermion model,
selective gapping occurs at hot spots indicated by the intersection of the 
Fermi surface arcs with the reduced diamond-shaped AF Brillouin zone\cite{schmalian} whereas in the spin-bag model\cite{bag}, a gap occurs only along the $(\pi,\pi)$ direction.  Neither of these models, however, possesses the strong
correlations intrinsic to the doped Mott state.

To conclude, our proposal that an orthogonality between the singlet and triplet states necessarily requires a finite length scale over which transport is governed by the distance between double occupancies implies 
that $U\rightarrow\infty$ and $L\rightarrow\infty$ do not commute. The emergence of such a finite length scale in the transport properties offers a possible resolution of the breakdown of the one-parameter scaling picture\cite{one}
for quantum criticality in the cuprates.  Finally, experiments\cite{diamag} demonstrating that diamagnetism in the pseudogap phase does not persist all the way to $T^\ast$ proves that
pair fluctuations alone cannot account for the pseudogap.  As advocated here, the pseudogap arises from Mottness and any relationship between ordering\cite{yazdani} or pairing and the pseudogap
is one of supervenience.

\acknowledgements We thank the NSF, Grant No. DMR-0305864, Duncan Haldane for a key discussion on the $U\rightarrow\infty$ and $L\rightarrow\infty$ limits and also T. Stanescu and A. Yazdani.


\begin{thebibliography}{99}
\bibitem{appears}There is no experimental evidence that the pseudogap is anisotropic at $T^\ast$.
\end{thebibliography}
\end{document}